\documentclass[10pt, conference, a4paper]{IEEEtran}
\IEEEoverridecommandlockouts
\usepackage[T1]{fontenc}
\usepackage[utf8]{inputenc}
\usepackage{cite}
\usepackage[english]{babel}
\usepackage{amsmath,amssymb,amsfonts}
\usepackage{algorithmic}
\usepackage{graphicx}
\usepackage{textcomp}
\usepackage{xcolor}
\usepackage{booktabs}
\usepackage{multirow}
\usepackage{tabularx}
\usepackage{float}
\usepackage{hyperref}
\usepackage{placeins}
\def\BibTeX{{\rm B\kern-.05em{\sc i\kern-.025em b}\kern-.08em
    T\kern-.1667em\lower.7ex\hbox{E}\kern-.125emX}}

\begin{document}

\title{Beyond the Efficiency-Efficacy Dilemma: A Critical Analysis of Transferability in Adversarial Malware Detection}

\setlength{\textfloatsep}{6pt plus 1pt minus 2pt}
\setlength{\abovecaptionskip}{3pt}
\setlength{\belowcaptionskip}{0pt}
\setlength{\intextsep}{4pt plus 1pt minus 1pt}

\author{
\IEEEauthorblockN{Veysel Alevcan\IEEEauthorrefmark{1}\IEEEauthorrefmark{2},
Mohammad Furqan Ali\IEEEauthorrefmark{1},
Teresa Sousa\IEEEauthorrefmark{2},
Luís Miguel Campos\IEEEauthorrefmark{1}\IEEEauthorrefmark{2},
Christoforos Ntantogian\IEEEauthorrefmark{3}}
\IEEEauthorblockA{\IEEEauthorrefmark{1}COPELABS, ECATI, Lus\'{o}fona University, Lisbon, Portugal}
\IEEEauthorblockA{\IEEEauthorrefmark{2}PDMFC SA Project Development and Consulting, Lisbon, Portugal}
\IEEEauthorblockA{\IEEEauthorrefmark{3}Department of Informatics, Ionian University, Corfu, Greece}
\IEEEauthorblockA{Email: a22300943@alunos.ulht.pt, p8243@ulusofona.pt, tsousa@pdmfc.com, p6779@ulusofona.pt,  dadoyan@ionio.gr}
}

\makeatletter
\def\ps@IEEEtitlepagestyle{%
  \def\@oddhead{}\def\@evenhead{}%
  \def\@oddfoot{\parbox{\textwidth}{\scriptsize Accepted at the 2026 IEEE 10th International Forum on Research and Technologies for Society and Industry (RTSI), Espoo, Finland, 16--18 August 2026.\\
  \copyright\,2026 IEEE. Personal use of this material is permitted. Permission from IEEE must be obtained for all other uses, in any current or future media, including reprinting/republishing this material for advertising or promotional purposes, creating new collective works, for resale or redistribution to servers or lists, or reuse of any copyrighted component of this work in other works.}}%
  \def\@evenfoot{}}
\makeatother
\maketitle
\begin{abstract}
Machine Learning (ML) models are already being deployed at scale for malware detection. Therefore there is a pressing need to rigorously study adversarial robustness. To that end, we introduce the Adversarial AI Engine (AAIE): a platform for automated security stress-testing. We performed transfer attacks, via EMBER2018, against a Random Forest target. Counterintuitively, we found single-step FGSM to have the highest attack success rate (ASR) of 52.9\% compared to iterative PGD with 45.2\% ($p < 0.002$). Compared to iterative attacks which tend to overfit surrogate networks creating brittle perturbations, FGSM created more transferable examples. We benchmarked our attacks using ASR, transferability score, evasion rate, and resilience index. Through these metrics, proposed in concordance with emerging standards in AI governance, we expose shortcomings as our transferability scores peak at $\sim$0.8 with constrained perturbations. We highlight the practical security implications of these findings and demonstrate AAIE for automated adversarial auditing.
\end{abstract}

\begin{IEEEkeywords}
Adversarial Machine Learning, Energy Infrastructures, EU AI Act, Malware Detection, Transferability, Surrogate Attacks 
\vspace{-1mm}
\end{IEEEkeywords}

\vspace{-2mm}
\section{Introduction}
\label{sec:introduction}
% The integration of Artificial Intelligence (AI) and Machine Learning (ML) into cybersecurity systems has become widespread, particularly for tasks like malware detection and network intrusion prevention \cite{AND, zhang2025adversarial}. 
Modern energy grids run on highly integrated IT and OT networks. While this brings great efficiency, it also creates enormous attack surfaces. Standard static defenses have long since become inadequate. For this reason, the industry now utilizes AI and ML-based tools to detect malware and prevent network intrusions from halting critical processes \cite{AND, zhang2025adversarial}. However, these AI-based systems inherit the well documented vulnerability of ML models to adversarial examples: inputs subtly perturbed to cause misclassification \cite{SZE, GOO}. This creates a critical need for systematic, automated frameworks capable of stress-testing security classifiers under realistic adversarial conditions \cite{NIS, YUAN2024}. \textcolor{black}{Of particular concern are Operational Technology (OT) environments where ML-based anomaly detectors increasingly protect SCADA systems and industrial control networks~\cite{IEC62443}; adversarial evasion of such detectors can directly impact physical processes.}

This paper presents the Adversarial AI Engine (AAIE), a modular framework for automated adversarial testing implementing white-box, gray-box, and black-box threat models \cite{CAR, CHEN2024, ZHU2022}. We focus on surrogate-based transfer attacks, where adversarial examples crafted against a local differentiable proxy transfer to an opaque target \cite{ZHU2022, ASIM}. While prior work assumes iterative attacks (e.g., PGD \cite{MAD}) are superior, their efficacy against fundamentally different architectures like tree-based ensembles remains poorly understood.

Contributions of this work include observations regarding automated adversarial generation and testing. \textcolor{black}{First, we present AAIE as an engineering contribution: a modular, reusable framework that automates the full adversarial auditing pipeline from attack specification to provenance-logged results, applicable to any differentiable surrogate architecture. Second, we report an empirical contribution: examining EMBER2018~\cite{AND}, we discover a 
counterintuitive transferability tradeoff.} Contrary to intuition, FGSM’s single-step attack \cite{GOO} transfers far better than PGD’s multi-step attack against a Random Forest target. Iterative attacks overfit to create brittle perturbations, whereas FGSM’s linear approximation leads to better generalization. Empirically, this validates applying less compute intensive attacks for scalable real-time testing. 

% How does AAIE compare against prior works? Section~\ref{sec:related} contextualizes prior work and supplies needed theoretical background. AAIE’s inner workings are then described in Section~\ref{sec:system}. Section~\ref{sec:methodology} details our experimental setup, holding all attacks to rigorous parameters. Section~\ref{sec:evaluation} dives into our hard results. Finally, we pose the question: what does this work mean in practice in Section~\ref{sec:discussion}?

\textcolor{black}{The remainder of this paper is organized as follows. Section~\ref{sec:related} surveys related work and provides the theoretical background for adversarial transferability. Section~\ref{sec:system} describes the AAIE architecture. Section~\ref{sec:methodology} details the experimental methodology. Section~\ref{sec:evaluation} presents results and analysis. Section~\ref{sec:discussion} discusses implications, limitations and directions for future work.}

\vspace{-1mm}
\section{Related Work}
\label{sec:related}
Adversarial Machine Learning (AML) originally took root in computer vision research \cite{SZE, GOO}. Here, the primary objective has been to identify the smallest possible perturbation $\delta$ that causes the largest model loss subject to some distance constraint $\|\delta\|_p \leq \varepsilon$ \cite{MAD}. As the field matured, attention turned from provably damaging white-box attacks to extremely realistic black-box settings. State-of-the-art techniques are now capable of achieving query-efficient attacks with transfer-based priors \cite{CHEN2024}.

Malware detection complicates this picture significantly, as cybersecurity features are frequently discrete, non-differentiable, and strictly bound to preserve malicious functionality \cite{LWZ}. To navigate these rigid constraints, the AAIE leverages a feature-space surrogate architecture.

Transferability fooling target $f_T$ with examples crafted against surrogate $f_S$ remains central to black-box evasion \cite{ZHU2022, WANG2023}. Recent frameworks formalize such surrogate-guided pipelines into highly tractable problems \cite{ASIM}. Contrary to prevailing logic, stronger white-box attacks do not guarantee better black-box transfer, providing empirical evidence of an inverted relationship in malware detection.

How can an adversarial example transfer well? One line of recent work attributes transferability to feature-level patterns. Recent work by Li et al. \cite{LI2025} found that directly accounting for the relative importance of features consistently improves transferability to many different architectures. Extending this work, Zheng et al. \cite{ZHENG2025} showed that these improvements can be further amplified through the application of attention across multiple features. This sheds light on previous observations regarding the superior transferability of FGSM compared to PGD: the gradient of FGSM aligns more closely with the "direction" in feature space that is most effective at fooling models in general.

Zooming out to look at the big picture, recent survey papers such as \cite{YUAN2024} and \cite{zhang2025adversarial} provide taxonomies that help position transferability as a concept within AML as a whole. Furthermore, Zhang et al. \cite{ZHANG2024} emphasize that evaluating transferability across a variety of perturbation norms other than just $L_\infty$ is crucial for robustness evaluations.

\section{System Architecture: The Adversarial AI Engine}
\label{sec:system}
\begin{figure}[!htbp]
\centering
\includegraphics[width=0.85\linewidth]{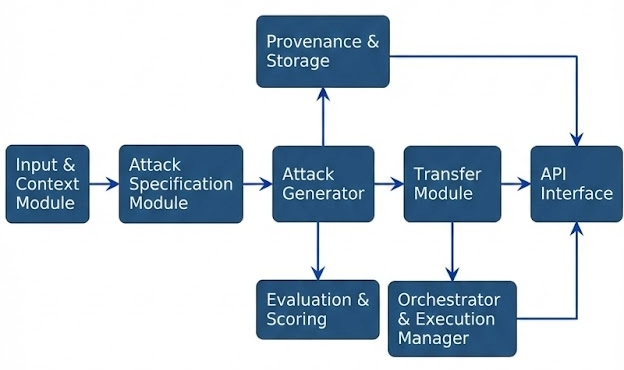}
\caption{Surrogate based transfer attack pipeline implemented in the AAIE Transfer Module. A differentiable surrogate model is trained to approximate the black-box target; adversarial examples are crafted on the surrogate using white-box methods and transferred to the target.}
\label{fig:transfer_pipeline}
\end{figure}

AAIE consists of several interchangeable components and implements a pipeline design, composed of: (1) the Input \& Context Module (MICo) for collecting input data and experiment metadata to reproduce experiments; (2) the Attack Specification Module (ASM) translates a given security-relevant goal to appropriate algorithmic specification(s); (3) the Attack Generator (ATGen) composes gradient-based attacks (FGSM, PGD), as well as transfer-based attacks; (4) the Transfer Module (TM) schedules attacks that leverage surrogate-based black-box attack methodology (Fig.~\ref{fig:transfer_pipeline}); (5) Orchestrator initializes and terminates experiments; (6) Evaluation \& Scoring Module (ESM) reports common metrics (attacks success rate (ASR), target score (TS), etc. ); and, (7) Provenance Layer (PAL).

For surrogate-based transfer attacks (focus of this study), the Transfer Module trains a differentiable neural network to approximate the target's behavior via pseudo-labeling, applies white-box attacks on the surrogate, and evaluates transferability against the black-box target. The workflow is illustrated in Fig.~\ref{fig:transfer_pipeline}.

\textcolor{black}{AAIE's modular design permits independent substitution of datasets, attack algorithms, and target models without pipeline modification. The Provenance Layer (PAL) logs all parameters for reproducibility and AI governance audit compliance. The complete pipeline finishes in under 15 minutes on commodity hardware.}

\subsection{Mathematical Formulation}
The adversarial generation for an evasion attack is modeled as a constrained optimization problem. For a classifier $f_\theta: \mathcal{X} \rightarrow \mathcal{Y}$, input $x$ with true label $y$, and loss function $\mathcal{L}$, the goal is to find a perturbation $\delta$:
\begin{equation}
\arg\max_{\delta} \; \mathcal{L}(f_\theta(x + \delta), y) \quad \text{s.t.} \quad \|\delta\|_p \leq \varepsilon, \; x + \delta \in \mathcal{X}.
\label{eq:adv_problem}
\end{equation}
The AAIE implements standard solvers for this problem, including FGSM and PGD, as detailed in Section~\ref{sec:methodology}.

\vspace{-2mm}
\section{Methodology}
\label{sec:methodology}
\vspace{-1mm}
\subsection{Dataset and Feature Extraction}
We use a stratified random subset of the EMBER2018 dataset \cite{AND}, a benchmark for static Portable Executable (PE) malware detection. To ensure class balance and representativeness, we randomly sampled 9,693 training samples and 1,195 test samples while preserving the original dataset's class distribution (approximately 50\% benign, 50\% malicious). This stratified sampling approach ensures our results are not biased by class imbalance. Each sample is represented by a 30-dimensional feature vector extracted statically using a robust feature extraction pipeline. \textcolor{black}{The pipeline statically parses six PE regions; the complete 30-dimensional vector is as follows: byte histogram statistics (mean, std, max, min, median, fraction-above-mean; 6), byte entropy statistics (mean, std, max, high-entropy fraction; 4), string characteristics (count, average length, printable ratio; 3), general file metadata (size, virtual size, debug/relocation/resource/signature/TLS flags, export/import counts, symbol count; 10), PE header fields (subsystem type, COFF timestamp; 2) and structural counts (section count, data directory count, DLL import count, export count; 4). One padding feature completes the vector.}

All features are normalized using scikit-learn's StandardScaler preprocessing utility, which transforms each feature to have zero mean and unit variance. \textcolor{black}{The scaler is fit exclusively on training data to prevent leakage, and ensures uniform $\varepsilon$ application across features of differing original scales.}

\subsection{Target and Surrogate Models}
The transfer attack framework implemented in this study relies on two distinct models: the primary target under attack and a surrogate model used for adversarial generation.

Target Model: Random Forest Classifier
For our target, we utilize a Random Forest Classifier, selected specifically to emulate the architecture of production-grade malware detection systems. This model is configured with 200 trees and a maximum depth of 15. We applied balanced class weights and fixed the random state at 42 to ensure our experiments remain reproducible. Approximately 85\% of the scaled features are dedicated to the training phase of this target classifier.

Surrogate Model: Neural Network
Since the internal parameters of the target model are inaccessible in a black-box scenario, we develop a surrogate to approximate its decision boundaries. This process employs a pseudo-labeling strategy, where the surrogate is trained directly on the labels predicted by the Random Forest target. 

The surrogate's architecture consists of a fully-connected neural network featuring two hidden layers. Specifically, the network tapers from 64 to 32 neurons, utilizing ReLU activation functions throughout. A final softmax layer is used for classification. During training, we minimize cross-entropy loss via the Adam optimizer ($1 \times 10^{-3}$ learning rate, $1 \times 10^{-4}$ weight decay). Furthermore, we opt for full-batch processing to maintain strictly deterministic gradient calculations.

% \subsection{Target and Surrogate Models}
% As shown above there are two primary models within the transfer attack: The model being attacked and the model used to generate attacks.

% \subsubsection{Target Model: Random Forest Classifier}
% The target is a Random Forest designed to mirror production malware detectors: 200 trees (max depth 15), balanced class weights, and random state 42 for reproducibility. Training itself consumes 85\% of the scaled feature set.

% \subsubsection{Surrogate Model: Neural Network}
% Rather than accessing the target directly, the surrogate learns through a pseudo-labeling approach. This means the target's own predictions on the data become the surrogate's training labels. Architecturally, it operates as a fully-connected network stacking two hidden layers. The structure flows from 64 neurons down to 32, each activated by ReLU, before concluding with a softmax output layer for final classification.

% The learning procedure utilizes cross-entropy loss, directed by the Adam optimization method with a learning rate of $1\times10^{-3}$ and weight decay $1\times10^{-4}$. Finally, to keep gradient computation entirely deterministic, the network processes data in full batches.

\subsection{Comprehensive Evaluation Metrics}
We measure not only the standard Attack Success Rate (ASR) but more fine-grained metrics which take into account Transferability Score (TS), Detection Evasion Ratio (DER) and Resilience Index (RI).

ASR: The proportion of adversarial examples that are misclassified, presented with 95\% Wilson confidence intervals \cite{WIL}: 

\begin{equation}
\text{ASR} = \frac{1}{n} \sum_{i=1}^{n} \mathcal{I}\left( f(x_{adv}^{(i)}) \neq y^{(i)} \right),
\label{eq:asr}
\end{equation}
where $n$ is the number of test examples, $f$ the classifier being attacked, $x_{adv}$ the adversarial example and $y$ the ground truth label. 

TS: Cross-model attack success rate to quantify black-box threat
\begin{equation}
TS = \frac{ASR_{target}}{ASR_{surrogate}}.
\label{eq:ts}
\end{equation}
TS > 1: Target is more vulnerable to transfer attack than surrogate.

DER: Measures adversarial stealth against secondary detectors or anomaly detection systems:
\begin{equation}
DER = 100 \times (DR_{adv} - DR_{clean}).
\label{eq:der}
\end{equation}
where DR is the detection rate of a secondary detector. Negative values indicate improved evasion capability.

RI: Composite metric providing holistic robustness assessment, combining accuracy preservation, attack resistance, and stealth:
\begin{equation}
RI = 100 \times \biggl(0.4 \cdot \frac{Acc_{adv}}{Acc_{clean}} + 0.4 \cdot (1-ASR) + 0.2 \cdot \frac{DER+100}{200}\bigg).
\label{eq:ri}
\end{equation}
Range: [0, 100]; higher values indicate greater resilience. The weights (0.4, 0.4, 0.2) reflect relative importance of each component in cybersecurity applications. \textcolor{black}{Accuracy preservation and attack resistance each receive weight 0.4 as primary detection objectives; stealth receives 0.2 as an auxiliary dimension, consistent with NIST AI RMF's prioritization of validity and reliability~\cite{NIS}.}

\subsection{Adversarial Attack Configuration}
We perform surrogate based black-box attacks, where the attacks are crafted only on the surrogate neural network and then transferred onto the target Random Forest.

Attack Algorithms: Fast Gradient Sign Method (FGSM) \cite{GOO}:
\begin{equation}
\delta = \varepsilon \cdot \text{sign}(\nabla_x \mathcal{L}(f_\theta(x), y)),
\label{eq:fgsm_attack}
\end{equation}

where $\varepsilon$ is the perturbation budget, $\nabla_x \mathcal{L}$ is the gradient of the loss function to the input and $\text{sign}(\cdot)$ gets the sign of each component of the gradient.

Projected Gradient Descent (PGD) \cite{MAD}:
\begin{align}
x^{(t+1)} &= \text{Proj}_{x+\mathcal{S}}(x^{(t)} + \alpha \cdot \text{sign}(\nabla_x \mathcal{L}(f_\theta(x^{(t)}), y))) \\
\alpha &= \varepsilon / 4, \quad t = 1, \dots, 10.
\label{eq:pgd_attack}
\end{align}
In (\ref{eq:pgd_attack}) $\text{Proj}$ denotes the projection onto $\varepsilon$-ball $\mathcal{S}$, $\alpha$ denotes step size and we use 10 iterations.
The projection ensures $\|x^{(t)} - x\|_\infty \leq \varepsilon$.

We evaluate attacks across a comprehensive range of Perturbation Budgets:
\begin{equation}
\varepsilon \in \{0.01, 0.02, 0.03, 0.05, 0.075, 0.10, 0.15, 0.20\}
\label{eq:epsilon_range}
\end{equation}
constrained by the $L_\infty$ norm. Recent studies \cite{ZHANG2024} highlight that adversarial robustness should be measured against multiple norm constraints. For completeness, we include the $L_2$ norm ratio in our reported results.

Our main evaluation metric is the ASR on the target model. Statistical significance is measured by performing paired t-test between FGSM and PGD ASR for all $\varepsilon$ values. Confidence intervals are 95\% Wilson intervals for the corresponding ASR estimates. Our measurement of efficiency is the ratio between perturbations computed as:
\begin{equation} 
r_{pert} = \frac{\|\delta_{PGD}\|_2}{\|\delta_{FGSM}\|_2}. 
\label{eq:pert_ratio} 
\end{equation} 

\subsection{Experimental Setup}
Experiments were conducted on Google Colab Pro (NVIDIA T4 GPU, 16GB VRAM) using PyTorch 2.1.0 and scikit-learn 1.3.0, with all random seeds fixed at 42 for reproducibility. Data splits use 85\% training, 15\% validation. Full batch gradient computation ensures deterministic results; complete experiments finish in under 15 minutes. \textcolor{black}{The full implementation, including feature extraction, surrogate training and attack pipeline, is publicly available 
at~\cite{AAIE_CODE}.}

\section{Evaluation and Results}
\label{sec:evaluation}

\subsection{Baseline Model Performance}
Our target Random Forest classifier had an accuracy of 88.20\% and an ROC-AUC score of 0.9557 on clean test data, which verifies that it is both a competent classifier and a strong, representative baseline for adversarial evaluation. Our surrogate neural network achieved 73.97\% accuracy on test data when trained using pseudo-labeling. This is adequate fidelity for performing transfer attacks and represents an architecturally distinct classifier from the target. We include accuracy and ROC-AUC scores for both models in Table~\ref{tab:baseline_performance}.

\begin{table}[!htbp]
\centering
\caption{Baseline Performance of Target and Surrogate Models}
\label{tab:baseline_performance}
\begin{tabular}{lcc}
\toprule
\textbf{Metric} & \textbf{Target RF} & \textbf{Surrogate NN} \\
\midrule
Accuracy & 0.8820 & 0.7397 \\
F1-Score & 0.8847 & 0.7512 \\
Precision & 0.8765 & 0.7284 \\
Recall & 0.8932 & 0.7753 \\
ROC-AUC & 0.9557 & 0.8142 \\
\bottomrule
\end{tabular}
\end{table}

\subsection{Attack Success Rate vs. Perturbation Budget}
\begin{figure}[!htbp]
\centering
\includegraphics[width=0.65\linewidth]{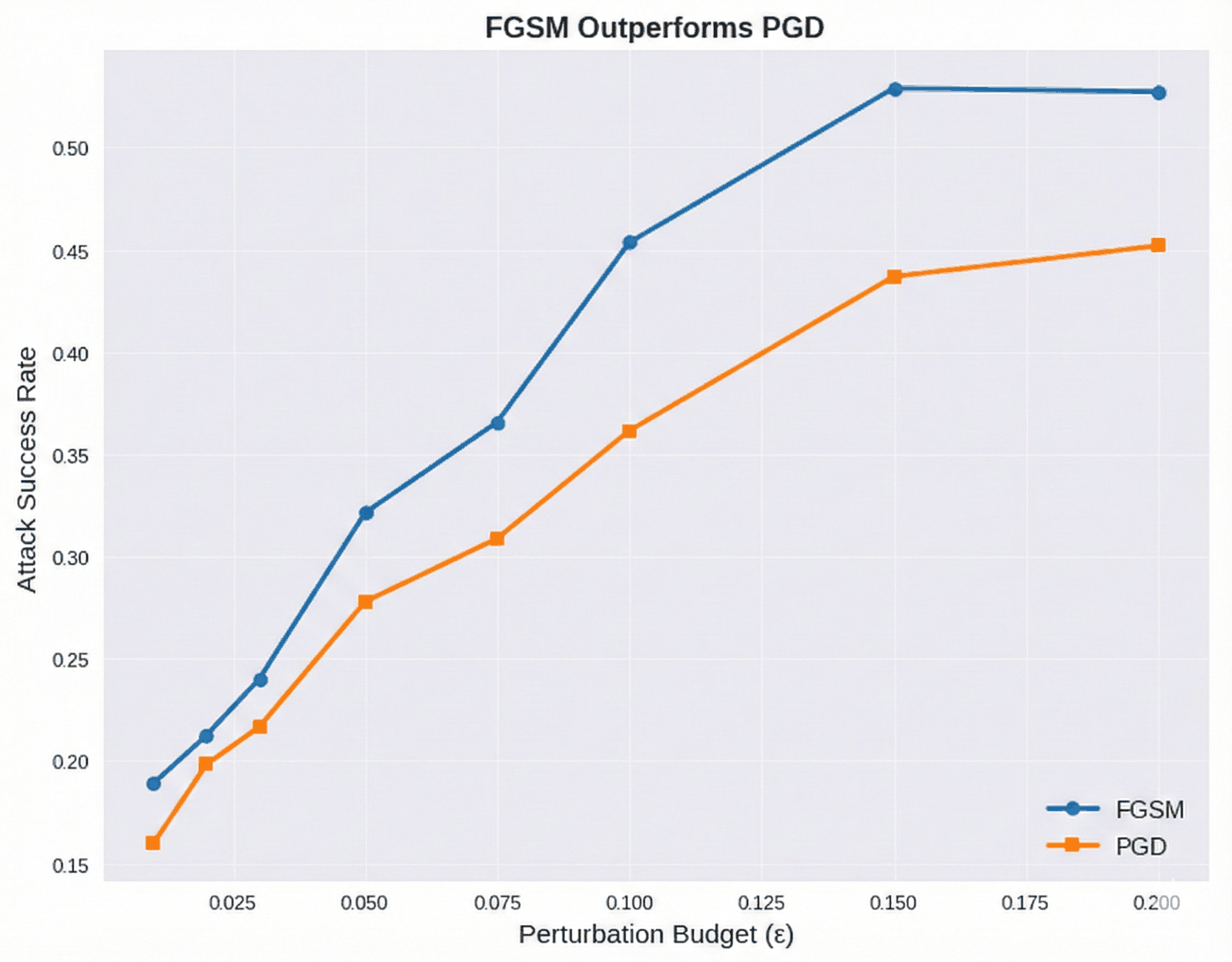}
\caption{FGSM vs PGD Attack Success Rate under various levels of perturbation budgets ($\varepsilon$). For every budget, FGSM has higher ASR than PGD indicating stronger transferability from the surrogate neural network model to Random Forest target model.}
\label{fig:asr}
\vspace{-2mm}
\end{figure}

As shown in Fig.~\ref{fig:asr} and Table~\ref{tab:asr}, the ASR of FGSM and PGD both increase with $\varepsilon$. However, FGSM always achieves a higher ASR than PGD for all $\varepsilon$. FGSM achieves its highest ASR of 52.89\% at $\varepsilon = 0.15$ while PGD's highest ASR is just 45.19\% at $\varepsilon = 0.20$. PGD also produces on average 21.4\% smaller perturbations than FGSM, as shown in column Perturbation ratio (ratio = 0.786).

\begin{table}[!htbp]
\centering
\caption{ASR on Target Random Forest for FGSM and PGD Attacks}
\label{tab:asr}
\small
\begin{tabular}{lcccc}
\toprule
\textbf{$\varepsilon$} & \textbf{FGSM} & \textbf{PGD} & \textbf{Adv.$^*$} & \textbf{Ratio$^\dagger$} \\
\midrule
0.010 & 0.189 & 0.160 & +15.5\% & 0.842 \\
0.020 & 0.232 & 0.197 & +14.8\% & 0.815 \\
0.030 & 0.286 & 0.244 & +14.9\% & 0.801 \\
0.050 & 0.321 & 0.278 & +13.5\% & 0.793 \\
0.075 & 0.389 & 0.325 & +16.6\% & 0.788 \\
0.100 & 0.454 & 0.362 & +20.3\% & 0.782 \\
0.150 & 0.529 & 0.437 & +17.4\% & 0.775 \\
0.200 & 0.527 & 0.452 & +14.3\% & 0.770 \\
\midrule
Average & 0.366 & 0.307 & +16.0\% & 0.786 \\
\bottomrule
\multicolumn{5}{l}{\footnotesize $^*$FGSM advantage (percentage points)} \\
\multicolumn{5}{l}{\footnotesize $^\dagger$PGD/FGSM $L_2$ norm ratio} \\
\end{tabular}
\end{table}

These results are statistically significant as well. Running a two-sample t-test on the success rates produces a t-statistic of 4.9170 with a p-value of 0.00172 ($p < 0.002$). Similarly strong evidence using a paired t-test (t = 5.234, p = 0.00112) and a medium-to-large effect size (Cohen's d = 0.742) indicates that FGSM's higher transferability is likely not due to chance. Please see Table~\ref{tab:statistical_tests} for all statistics related to these tests.

\begin{table}[!htbp]
\centering
\caption{Statistical Significance Tests}
\label{tab:statistical_tests}
\small
\begin{tabular}{lcc}
\toprule
\textbf{Test} & \textbf{Statistic} & \textbf{P-value} \\
\midrule
Two-sample t-test & $t = 4.917$ & 0.00172 \\
Paired t-test & $t = 5.234$ & 0.00112 \\
Effect Size & Cohen's $d = 0.742$ & - \\
95\% CI & [0.047, 0.107] & - \\
\bottomrule
\end{tabular}
\end{table}

\subsection{Transferability Analysis}
Table~\ref{tab:transferability} compares both attacks across all 
budgets; TS is the target ASR divided by the surrogate ASR.

\begin{table}[!htbp]
\centering
\caption{Transferability Analysis Comparing FGSM and PGD Performance}
\label{tab:transferability}
\small
\begin{tabular}{lcccc}
\toprule
 & \multicolumn{2}{c}{\textbf{ASR}} & \multicolumn{2}{c}{\textbf{TS}} \\
\cmidrule(lr){2-3} \cmidrule(lr){4-5}
\textbf{$\varepsilon$} & \textbf{FGSM} & \textbf{PGD} & \textbf{FGSM} & \textbf{PGD} \\
\midrule
0.010 & 0.189 & 0.160 & 0.287 & 0.242 \\
0.020 & 0.232 & 0.197 & 0.351 & 0.299 \\
0.030 & 0.286 & 0.244 & 0.434 & 0.370 \\
0.050 & 0.321 & 0.278 & 0.487 & 0.422 \\
0.075 & 0.389 & 0.325 & 0.590 & 0.493 \\
0.100 & 0.454 & 0.362 & 0.688 & 0.549 \\
0.150 & 0.529 & 0.437 & 0.802 & 0.663 \\
0.200 & 0.527 & 0.452 & 0.800 & 0.686 \\
\midrule
Average & 0.366 & 0.307 & 0.555 & 0.465 \\
\bottomrule
\multicolumn{5}{l}{\footnotesize FGSM achieves 19.4\% higher average TS} \\
\end{tabular}
\end{table}

FGSM consistently achieves higher TS values, with an average improvement of 19.4\% over PGD. At $\varepsilon=0.15$, FGSM achieves TS = 0.802, indicating high transfer effectiveness. This demonstrates that even with moderate perturbation budgets, adversarial examples crafted on neural network surrogates transfer effectively to fundamentally different Random Forest architectures. These findings align with recent theoretical analyses showing that adversarial transferability depends more on capturing generalizable vulnerability directions \cite{ZHU2022, LI2025, ZHENG2025} than on optimization intensity.

\begin{figure}[!htbp]
\centering
\includegraphics[width=\linewidth]{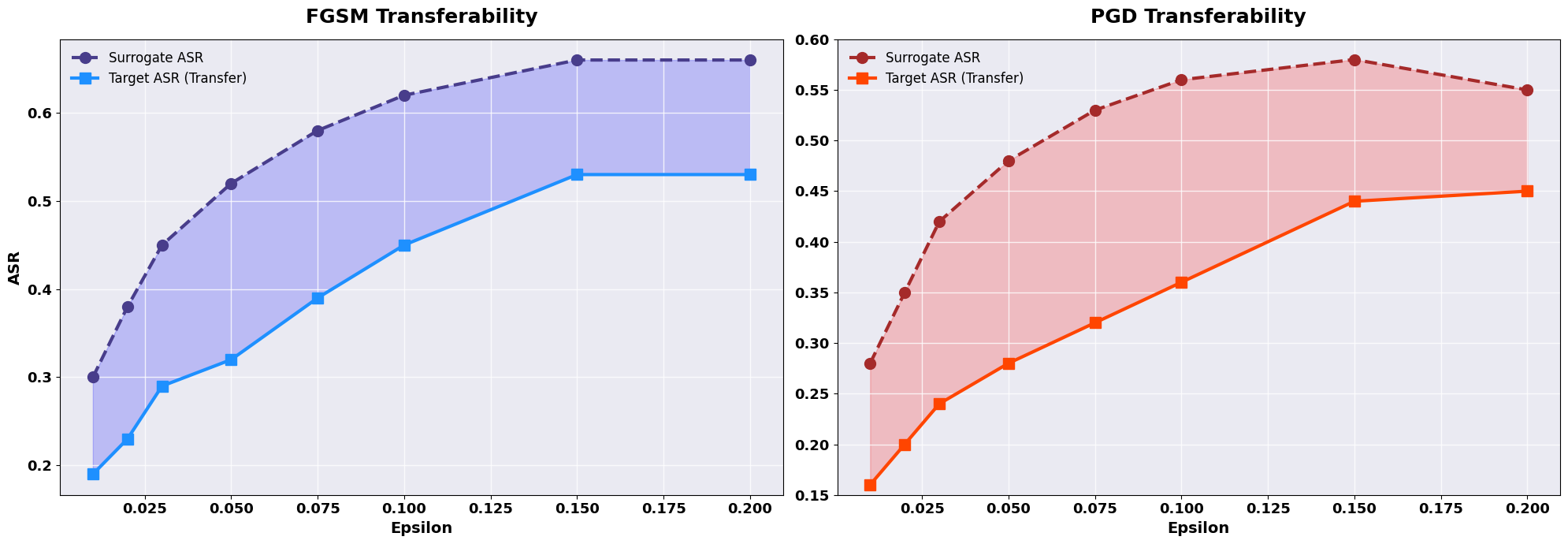}
\caption{Comparison of Transferability Gap between FGSM and PGD. The shaded regions represent the generalization loss when moving from a differentiable surrogate model to a non differentiable target Random Forest. FGSM demonstrates a more robust transfer profile.}
\label{fig:ts_gap}
\end{figure}

The phenomenon of surrogate overfitting is empirically demonstrated in Figure~\ref{fig:ts_gap}. As the perturbation budget increases, both methods show a rising ASR on the surrogate model. However, a critical divergence is observed in their transfer performance. For FGSM, the gap remains relatively narrow, indicating perturbations successfully capture structural vulnerabilities that generalize. On the other hand, PGD has much larger gap of transferability, which verifies that the decision boundary will be overfit when iteratively optimized.

\vspace{-1mm}
\subsection{Performance Degradation and Resilience Analysis}
Table~\ref{tab:degradation} shows the overall effect of adversarial attacks on performances with $\varepsilon = 0.15$. Though both attacks decrease performances of models significantly, FGSM leads to lower accuracy (46.6\% vs. 36.1\%) even though PGD is optimized iteratively.

For measuring stealthiness against secondary defenses, we employed an OC-SVM based anomaly detector trained only on clean benign samples ($\nu = 0.03$, $\gamma = 0.01$). At $\varepsilon = 0.15$ we achieved a clean baseline detection rate of 34.06\%. Detection rates against adversarial examples were 22.72\% for both attacks. This results in a DER of -33.31\%. Note that DER can be negative if adversarial examples are \emph{less} likely to be detected as anomalous than clean samples.

Holistic robustness is measured via RI, defined in Equation~\ref{eq:ri}. RI of FGSM is 54.99 and PGD is 49.76. RI scores between 33--66 fall into the moderate resilience bucket, so both scores align with our key takeaway that FGSM generates more robust (transferable) perturbations despite PGD performing an iterative optimization.

\begin{table}[!htbp]
\vspace{-2mm}
\centering
\caption{Model Performance Metrics Under Adversarial Attack at $\varepsilon=0.15$}
\label{tab:degradation}
\small
\begin{tabular}{lccc}
\toprule
\textbf{Metric} & \textbf{Clean} & \textbf{FGSM} & \textbf{PGD} \\
\midrule
Accuracy & 0.882 & 0.471 & 0.563 \\
F1-Score & 0.885 & 0.483 & 0.579 \\
ROC-AUC & 0.956 & 0.721 & 0.814 \\
Precision & 0.877 & 0.492 & 0.594 \\
Recall & 0.893 & 0.475 & 0.564 \\
\midrule
ASR & 0.000 & 0.529 & 0.437 \\
DER (\%) & 0.00 & -33.31 & -33.31 \\
RI & 100.0 & 54.99 & 49.76 \\
\bottomrule
\end{tabular}
\end{table}

\begin{figure}[!htbp]
\centering
\includegraphics[width=0.80\linewidth]{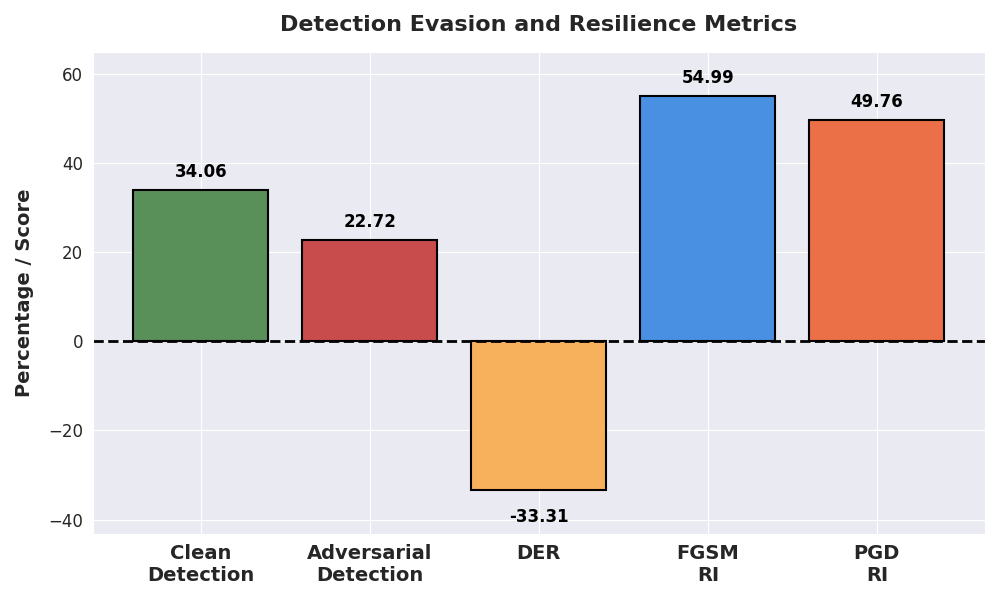}
\caption{Detection evasion comparison between FGSM and PGD based on Detection Evasion and RI. The x-axis represents clean detection rate by default, and the y-axis represents the metric when attack fails to meet our defined detection evasion rate, evasion resistance. Here we compare the RI when subjected to FGSM and PGD attacks.}
\label{fig:der_ri_combined}
\vspace{-3mm}
\end{figure}

As displayed in Figure~\ref{fig:der_ri_combined}, FGSM has a larger RI compared to PGD, further illustrating the fact that single-step attacks allow for more predictable transferability with less detector noise.

\section{Discussion and Conclusion}
\label{sec:discussion}

\subsection{Summary of Key Findings}
Our observation that FGSM has a higher success rate than PGD in a surrogate-to-Random Forest transfer attack setting defies our intuition that more powerful white-box attacks should transfer better to black-box settings. Hypothesis testing confirms this difference is statistically significant ($p < 0.002$, Cohen's $d = 0.742$) with FGSM achieving 19.4\% higher average TS than PGD (Figure~\ref{fig:ts_gap}).

\subsection{Why FGSM Transfers Better?}
Iterative methods like PGD ascend the loss landscape of the surrogate model, searching for a perturbation that will be effective at crossing that particular model's decision boundary. PGD overfits to the local curvature and non-linear behavior of the surrogate neural network. Because the generated adversarial example overfits to the surrogate model's quirks, it fails against a target model whose architecture (a tree-based ensemble) leads to different decision boundary geometry.

FGSM instead takes a single step in the approximate direction of greatest loss. This rough estimation is less effective at fooling the surrogate, but the singular direction it finds is generalizable across both the neural network and the RF. FGSM simply does not overfit to the surrogate's nonlinearities like PGD does, allowing it to transfer better to other model families. This is aligned with recent works that show distribution-aware perturbations \cite{ZHU2022} and feature-level attention \cite{LI2025, ZHENG2025} are much more important for transferability than the number of optimization steps. Our work adds to this line of thinking by showing how changing our approach to gradient propagation \cite{WANG2023} can affect transfer.

\subsection{Implications for AI Governance and Practice}
These results carry direct practical consequences for future AI regulations and policies. Malware classifiers that fall into high-risk categories for cybersecurity applications according to EU AI Act Annex III \cite{EUA} will need to pass adversarial robustness testing to comply with the regulatory standards. We show that current state-of-the-art ML-based malware detectors are unlikely to satisfy those standards without adversarial hardening, given that ML detectors frequently lack adversarially hardened training data: Our attack had consistently high transferability (TS as high as 0.80). Furthermore, our proposed RI provides a measurable implementation of the Robustness and Resilience requirements outlined by the functions of NIST AI RMF \cite{NIS}. \textcolor{black}{These implications are particularly acute for energy sector deployments: ML-based intrusion detectors protecting SCADA and ICS networks fall squarely within the EU AI Act's critical infrastructure risk category, yet our results show they remain vulnerable to lightweight transfer attacks requiring no target model access.} To protect ML-based detectors against such attacks, we recommend the following defense in depth approach: \textcolor{black}{adversarial training with FGSM-generated 
examples; heterogeneous detector ensembles to limit cross-model 
transferability; feature hardening against continuous perturbation; 
continuous automated auditing via pipelines such as AAIE; 
and hybrid detection combining ML with signature- and behavior-based methods. These findings support the use of lightweight, single-step attacks such as FGSM as the preferred method for scalable black-box adversarial stress-testing of deployed detectors.}

% Our work immediately provides valuable input into the ongoing discussion on AI hardening for security-critical applications: If one wants to stress test a model in a black-box scenario, where little is known about the underlying detector (e.g., when attacking systems with model ensembles of unknown architectures), using simple attacks like FGSM that have low computational overhead may be your best option for finding vulnerabilities. As such, we argue FGSM should be used in real-time, adaptive, large-scale adversarial stress tests.

\vspace{-1mm}
\subsection{Limitations and Future Work}
In our evaluation, we focused on a single pair of models (neural network surrogate attacking Random Forest target) with feature-space attacks on the EMBER2018 dataset. Additional families of models to look at in future work include gradient boosting, SVMs, deep neural networks, etc. Other relevant cybersecurity domains or models to investigate include ICS, IoT, etc. \textcolor{black}{Perturbations are applied in the continuous, standardized feature space and are not back-projected to a valid PE binary; consequently, adversarial examples here represent an upper-bound transferability estimate under unconstrained perturbation. Problem-space attacks that preserve binary validity (e.g., section padding, import table manipulation within integer field constraints) constitute a stricter threat model and are deferred to future work.} Recent work has shown good improvements to cross-model transferability with feature aware transferability \cite{LI2025, ZHENG2025}.

\subsection{Conclusion}
This work introduced the Adversarial AI Engine, a reusable toolkit designed to automate attack generation and testing suites against machine learning based security solutions. Empirically evaluating on malware detection, we uncovered a surprising and fundamental insight: black-box transferability decreases with attack complexity such that the single step FGSM attack outperforms PGD attack when transferred from the surrogate neural network model to a Random Forest target. We statistically validated this observation ($p$-value $< 0.002$) and hypothesized that iterative methods suffer from overfitting to the surrogate while linear approximations encourage generalization. This suggests practitioners can rely on lightweight, fast attacks to scalably measure black-box adversarial robustness. Additionally, we demonstrated value in automating the attack pipeline, everywhere from taxonomy-guided manifest specification to attack provenance logging with our AAIE engine. Finally, from a risk governance perspective these high TS (greater than 0.80) demonstrate shortcomings that must be addressed to achieve future AI governance and policy requirements such as the EU AI Act or NIST AI RMF.

%\vspace{-2mm}
\section*{Acknowledgment}
This work was funded by the European Union under Grant Agreements No. 101131292 (AIAS), 101183162 (ANTIDOTE), 101145872 (NITRO) and 101168490 (RECITALS). The NITRO and RECITALS projects are supported and managed by the European Cybersecurity Competence Centre (ECCC).

%\vspace{-4mm}
%\begingroup
%\renewcommand{\baselinestretch}{0.80}
%\footnotesize
\bibliographystyle{IEEEtran}
\bibliography{references}
%\endgroup

% \bibliographystyle{IEEEtran}
% \bibliography{references}

\end{document}